\documentclass[sigconf]{acmart}

\AtBeginDocument{%
  \providecommand\BibTeX{{%
    \normalfont B\kern-0.5em{\scshape i\kern-0.25em b}\kern-0.8em\TeX}}}

\setcopyright{acmcopyright}
\copyrightyear{2024}
\acmYear{2024}

\acmConference{ETRA ’24}{June 04--07,
  2024}{Glasgow, UK}

\begin{document}

\title{Enhancing Saliency Prediction in Monitoring Tasks: The Role of Visual Highlights}


\author{Zekun Wu}
\email{wuzekun@cs.uni-saarlaned.de}
\orcid{0000-0002-5233-2352}
\affiliation{%
  \institution{Saarland University}
  \streetaddress{Campus, 66123 Saarbrücken}
  \city{Saarbrücken}
  \state{Saarland}
  \country{Germany}
  \postcode{66123}
}

\author{Anna Maria Feit}
\email{feit@cs.uni-saarland.de}
\affiliation{%
  \institution{Saarland University, Saarland Informatics Campus}
  \streetaddress{E1 7, 66123 Saarbrücken}
  \city{Saarbrücken}
  \country{Germany}}

\renewcommand{\shortauthors}{Wu and Feit}

\begin{abstract}
This study examines the role of visual highlights in guiding user attention in drone monitoring tasks, employing a simulated interface for observation. The experiment results show that such highlights can significantly expedite the visual attention on the corresponding area. Based on this observation, we leverage both the temporal and spatial information in the highlight to develop a new saliency model: the highlight-informed saliency model (HISM),  to infer the visual attention change in the highlight condition. Our findings show the effectiveness of visual highlights in enhancing user attention and demonstrate the potential of incorporating these cues into saliency prediction models.  
\end{abstract}



\begin{teaserfigure}
  \centering
  \includegraphics[width=0.9\textwidth]{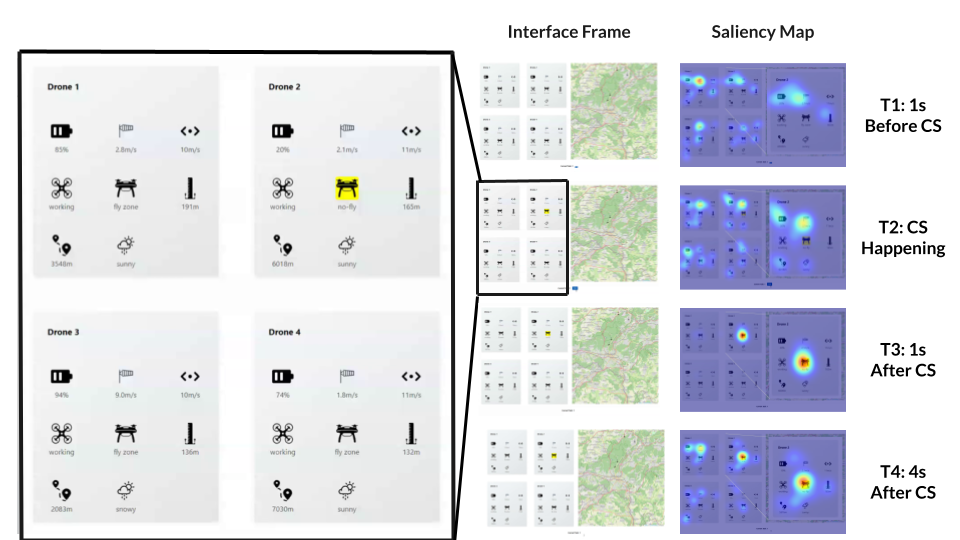}
  \caption{Visual highlights can effectively direct the user's attention to relevant areas. On average, it takes around 1s until the highlight is noticed and some user's attention has already moved away after 4s, as shown by the sequence of interface screenshots and saliency maps from the multi-drone monitoring task we studied in this work. The images show a critical situation (CS) occurring for the second drone and the corresponding saliency maps at four-time points: 1s before CS, CS happening, 1s after CS, and 4s after CS. Note that with the highlight's help, an immediate focus of attention would be put on the CS icon as visualized by the heatmap in the saliency map at T3 (1s after CS). Image-based saliency models, as typically employed to predict attention on user interfaces, cannot capture the temporal change in visual attention due to visual highlighting.}
  \Description{Interface Frame and Saliency Map in different Timestamp}
  \label{fig: screenshot}
\end{teaserfigure}

\maketitle


\label{Motivation}
\section{Motivation and Study Design}
In this study, we aim to explore how visual highlighting affects a user's visual attention during a monitoring task, and, more importantly, how we can \emph{predict} the temporal changes in visual attention due to highlighting, using saliency models. As indicated in  \autoref{fig: screenshot}, we developed an interface simulating the task of remotely monitoring four delivery drones. In the task, the participants had to identify critical situations (CS) that could potentially happen at any moment of the flight, while keeping an eye on all the drones' states to maintain the situation awareness (SA). When the CS happened, the corresponding icon would be highlighted in yellow with a 50\% probability. Participants were instructed to press the space bar when they detected a CS and at the same time monitor all parameters of all drones. Also, to test the participants' SA, we would stop the task randomly to ask multiple choice questions targeting the specific drone state (for example, what is the altitude for each drone now?), referring to the Situation Awareness Global Assessment Technique (SAGAT)~\cite{endsley2000direct}. We have recruited 30 participants and kept effective data from 28 of them. The study procedure and task were approved by the university’s ethics committee.



\section{Visual Highlights' Influence on Gaze Behavior}

\begin{figure}[h]
\centering
\includegraphics[width=\linewidth]{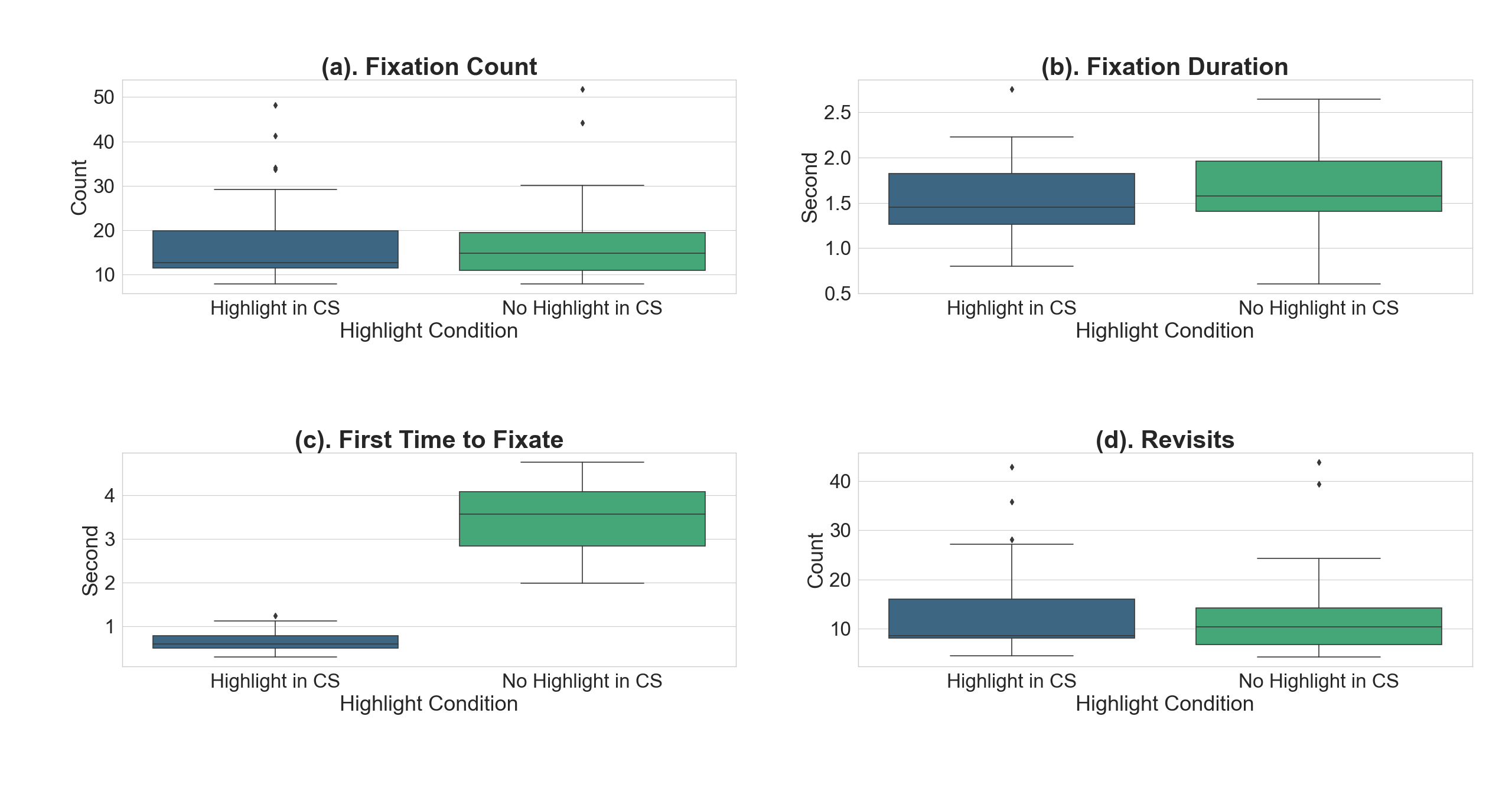}
\caption{Hit Rates and Response Time in Different Highlight Conditions}
\label{fig: Gaze}
\end{figure}

We started with gaze data analysis for different highlight conditions. Specifically, we chose the icon corresponding to CS and the parameter below it as an area of interest (AOI) and used a set of gaze metrics as fixation count, fixation duration, time to first fixation, and revisits based on prior work ~\cite{feit2017toward} to give insights into the highlights' influence on the user's visual attention. Our findings, illustrated in \autoref{fig: Gaze}, reveal no significant differences in fixation count and fixation duration between the two conditions. However, the time to first fixation was shorter for highlighted situations than non-highlighted ones, suggesting that visual highlighting effectively directs attention more rapidly in CS. The number of revisits to the AOI showed no significant difference between the two conditions, indicating that once a CS is detected, visual highlighting does not further influence visual attention in terms of fixation behavior. These findings are consistent with the previous studies ~\cite{karatas2020evaluation,schriver2017expertise} and confirm that visual cues can facilitate faster visual attention in the context of a monitoring task where the user has to retain high SA of all drones while also detecting CS.

\section{Highlight Informed Saliency Prediction}

\begin{figure}[h]
\centering
\includegraphics[width=\linewidth]{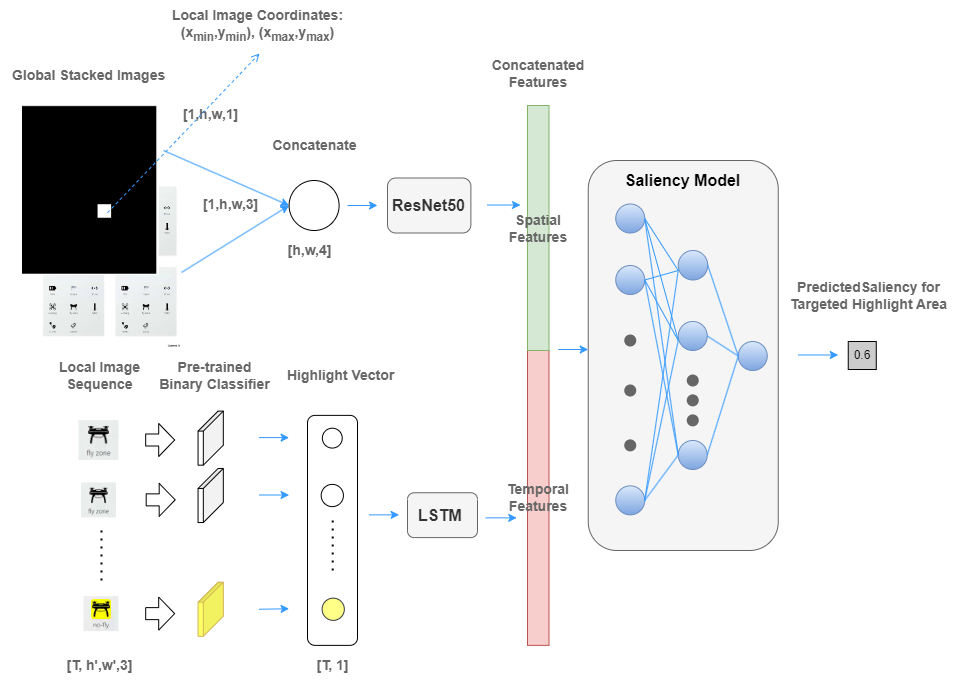}
\caption{Highlight Informed Saliency Model Framework, which includes three key components: the ResNet50 to extract the spatial information, the two-layer LSTM block to generate the temporal features, and the MLP takes both types of features to predict the saliency of the AO}
\label{fig: HISM}
\end{figure}

To predict how the visual highlight affects users' visual attention over time, we developed a new deep-learning highlight-informed saliency model (HISM), as indicated in \autoref{fig: HISM}. The HISM is targeted at predicting how salient the AOI is relative to other visual elements in different timestamps with different highlight conditions. Specifically, in the drone monitoring interface, we chose the icons and the parameters below them as the potential visual elements. We would normalize the saliency in each timestamp over all of the elements to calculate the element-level saliency in the temporal sense \cite{gupta2018saliency, aydemir2023tempsal}.

\begin{figure}[h]
\centering
\includegraphics[width=\linewidth]{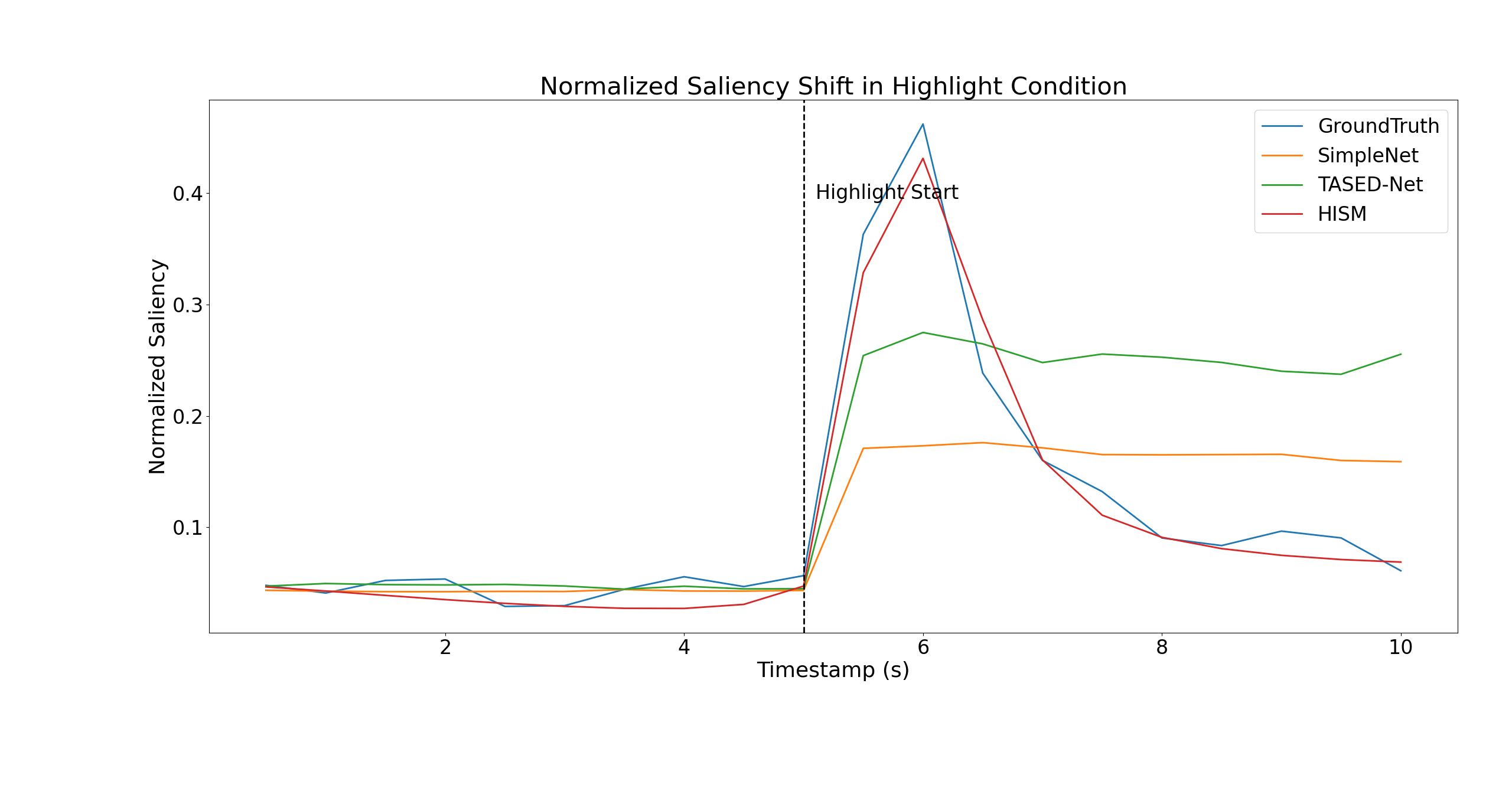}
\caption{Normalized saliency of the critical AOI over time computed over 0.5s windows. The corresponding icon was highlighted at second 5. Our HISM model captures the time to first fixation of about 1s and the following decrease in attention. State-of-the-art image- and video-based saliency models (SimpleNet and TASED-Net) could not capture the temporal changes in users' attention}
\label{fig: result}
\end{figure}

The HISM takes the input information from the visual highlight in two branches: the spatial information as the global stacked image that contains both the global picture of the interface screen and the exact location of the targeted AOI and the temporal information as a sequence of the local image of the targeted AOI in different timestamps. As indicated in the \autoref{fig: HISM}, the HISM mainly consists of three components: the ResNet\cite{he2016deep} extracting the spatial features from the global stacked image, the LSTM\cite{yu2019review} extracting the temporal features from the highlight vector, which is translated from the local image sequence with pre-trined binary classifier, and the Saliency Model with three fully connected layers to predict the final saliency. To examine the HISM performance, we chose two current state-of-art saliency model: SimpleNet\cite{reddy2020tidying} and TASED-Net \cite{min2019tased}, as baselines. As both of them are predicting saliency on the pixel level, we normalized their saliency prediction results over the AOI being highlighted. As indicated in \autoref{fig: result}, by considering both the temporal and spatial information of the highlight, the HISM outperforms both baselines as it exhibits enhanced predictive accuracy with a mean squared error (MSE) of 0.006, which is markedly lower than that of the TASED-Net and SimpleNet models, which achieved performance with MSEs of 0.01 and 0.009, respectively. 

\label{conclusions}
\section{Conclusions and Future Work}
In short, we developed a saliency model targeting predicting the visual attention change on the visual element with different highlights in the monitoring task. In the future, we will further examine the model performance on the free-viewing gaze data from more general interfaces such as web pages, etc. 

\begin{acks}
This work is funded by DFG grant 389792660 as part of TRR~248 -- CPEC, see \url{https://perspicuous-computing.science}
\end{acks}


\bibliographystyle{ACM-Reference-Format}
\bibliography{references}

\end{document}